\documentclass{bmcart}

\usepackage[utf8]{inputenc} 
\usepackage{amsmath}
\usepackage{tabu}
\usepackage{booktabs}
\usepackage{multirow}
\usepackage{graphicx}
\usepackage{wrapfig}
\usepackage{soul}
\usepackage{xurl}
\usepackage{enumitem}
\usepackage{xcolor,colortbl}
\usepackage{lineno}
\usepackage{setspace}

\startlocaldefs
\endlocaldefs

\def \fullwidth{0.9\columnwidth}
\begin{document}

\newcommand\tab[1][1cm]{\hspace*{#1}}
\newcommand{\coviddataset}{Twitter\_Covid}
\newcommand{\midtermdataset}{Twitter\_Midterm}
\newcommand{\fbdataset}{Facebook\_Midterm}
\newcommand{\embdiff}{node2vec\_diff}
\newcommand{\embtrust}{node2vec\_trust}
\newcommand{\colorbest}{\cellcolor{blue!30}}
\newcommand{\colorsecondbest}{\cellcolor{blue!10}}

\begin{frontmatter}

\begin{fmbox}
\dochead{Research}


\title{Account credibility inference based on news-sharing networks}


\author[
   addressref={aff1},                   
   corref={aff1},                       
   email={baotruon@iu.edu}   
]{\inits{BT} \fnm{Bao Tran} \snm{Truong}}
\author[
   addressref={aff1,aff2},
]{\inits{OMA} \fnm{Oliver Melbourne} \snm{Allen}}
\author[
   addressref={aff1},
]{\inits{FM} \fnm{Filippo} \snm{Menczer}}


\address[id=aff1]{
  \orgname{Observatory on Social Media, Indiana University}, 
  \city{Bloomington},                              
  \cny{US}               
}
\address[id=aff2]{%
  \orgname{Network Science Institute, Northeastern University},
  \city{Boston},
  \cny{US}
}



\end{fmbox}


\begin{abstractbox}

\begin{abstract} 
The spread of misinformation poses a threat to the social media ecosystem. Effective countermeasures to mitigate this threat require that social media platforms be able to accurately detect low-credibility accounts even before the content they share can be classified as misinformation. Here we present methods to infer account credibility from information diffusion patterns, in particular leveraging two networks: the reshare network, capturing an account's trust in other accounts, and the bipartite account-source network, capturing an account's trust in media sources.
We extend network centrality measures and graph embedding techniques, systematically comparing these algorithms on data from diverse contexts and social media platforms. 
We demonstrate that both kinds of trust networks provide useful signals for estimating account credibility.
Some of the proposed methods yield high accuracy, providing promising solutions to promote the dissemination of reliable information in online communities.
Two kinds of homophily emerge from our results: accounts tend to have similar credibility if they reshare each other's content or share content from similar sources. 
Our methodology invites further investigation into the relationship between accounts and news sources to better characterize misinformation spreaders.
\end{abstract}




\end{abstractbox}
%

\end{frontmatter}



\onehalfspacing
\section{Introduction}

Many people are now getting news from social media \cite{gottfried2016news}. With just a click on the ``Share'' button, anyone can be a broadcaster of news on these platforms. Such a low barrier, combined with uneven journalistic standards in online news, has facilitated the spread of misinformation in the information ecosystem \cite{zarocostas2020infodemic}. 
Such proliferation of misinformation poses grave threats to  democracy~\cite{woolley2018computational}, the economy~\cite{fisher2013syrian}, and public health~\cite{tasnim2020impact,allington2020health,pierri2022online,yang2021twitterfb}.

Existing methods to curb misinformation focus on classifying either the content or the account posting it. However, multiple challenges exist for content-based methods, posing a need for methods to evaluate sources instead of (or in addition to) the content itself.
In particular, traditional fact-checking methods involving human efforts to manually verify the accuracy of claims cannot scale with the sheer volume and speed of information being shared online. 
Automatic misinformation detection could potentially overcome the problem of scale. But when they work, these methods rely on extensive language features such as writing style, lexicon, and emotion \cite{zhou2020fake}. Therefore they need to be constantly retrained to reflect new knowledge and evolving tactics employed by purveyors of false information. 
These challenges are exacerbated by the rise of content created by generative AI. The availability of open-source large language models (LLMs) brings down the cost of generating content, creating opportunities for malicious actors to spread misleading content and influence public opinion \cite{goldstein2023generative,Menczer2023AI-harms}.
Worse yet, LLMs' ability to mimic writing styles makes this type of content persuasive yet very challenging to detect \cite{openai2023aiclassifier}. 

In this paper, we propose several methods to infer the credibility of news-sharing accounts on social media to detect low-credibility accounts likely to spread misinformation.  
The credibility of a news source or account may depend on many factors, including reputation and adherence to factual reporting and transparency. 
More precisely, we define \emph{high-credibility news sources} to be those that meet objective journalistic criteria as assessed by third-party fact-checking organizations \cite{hovland1951influence, westerman2014social}. Following that, \emph{credible accounts} are those who share or reshare high-credibility sources. 
Credibility indicators provide useful signals for consumers to navigate the information landscape: they not only help people seeking information outlets \cite{turcotte2015news} but can also decrease the propensity to share misinformation \cite{Yaqub2020} by raising user awareness. 
Knowing unreliable, influential accounts might also aid platforms in mitigating their impact. 

News-sharing decisions on social media depend on the actual content as well as on \textit{trust} in \textit{who} shares it \cite{turcotte2015news, mediainsight2017whoshared} and \textit{what} media outlet it originates from \cite{sterrett2019shared}.
When information sharing on social media is represented as a network connecting accounts and/or news sources, we can apply network analysis methods to infer node properties by propagating labels across links \cite{Mishra,rath2018utilizing,bild2015aggregate}. 
Most existing network-based methods to detect low-credibility accounts \cite{rath2018utilizing} focus solely on interactions between accounts. 
However, bipartite networks capturing the reinforcing relationship between news outlets and consumers have been shown to be effective for classifying misinformation content \cite{shu2019studying}. 
This merits further examination of the interactions between accounts and sources to better understand the characteristics of misinformation spreaders. 

Social context is useful in detecting fake news \cite{shu2019beyond}. Generalizing this observation, we hypothesize that it is possible to infer the credibility of an account by looking at either the sources or the other accounts they trust. 
The fundamental assumption underlying such inference is the existence of \textit{credibility homophily} among misinformation spreaders. Homophily can be defined based on different network relationships, such as following, resharing, co-sharing, and so on.
For example, adjacent nodes in the reshare network (accounts that reshare each other) trust each other, so they should have similar credibility. 
While \textit{trust in accounts} and \textit{trust in sources} are both reasonable assumptions for the task, no other work has compared their effectiveness in the same settings. 

We explore three research questions:
\begin{itemize}[label={}]
    \item \textbf{Q1:} Is there credibility homophily in the reshare network? 
    \item \textbf{Q2:} Is there credibility homophily in the co-share network? 
    \item \textbf{Q3:} Can we infer the credibility of accounts by leveraging such homophily and looking at neighbor nodes in the reshare or bipartite/co-share networks, i.e., by examining the trust relationships among accounts or between accounts and sources?    
\end{itemize}

In this paper, we propose several network-based methods --- including centrality measures and graph embeddings --- to infer the credibility of news-sharing accounts on social media. 
We explore \textbf{Q1} using the reshare network, where a directed edge represents trust by one account in another. We explore \textbf{Q2} using the bipartite network where accounts are connected to the news sources they share. 
The paper makes three contributions:

\begin{enumerate}
    \item We introduce several methods to measure the credibility of accounts by leveraging different kinds of information-sharing networks.
    
    \item We introduce an evaluation framework to 
    systematically estimate and compare the accuracy of our algorithms using empirical data from multiple contexts and social media platforms.
        
    \item We show that there are two kinds of homophily among information spreaders: accounts tend to reshare content from individuals with similar credibility (\textbf{Q1}) and to share content from the same sources as individuals with similar credibility (\textbf{Q2}). These diffusion patterns explain the effectiveness of network methods that estimate an account's credibility using their trust in other accounts or in sources (\textbf{Q3}).
\end{enumerate}

After reviewing related work, we present our methodology, including the definition of the credibility inference task and algorithms leveraging an account's trust in other accounts and in sources, respectively. We then describe the experimental setup and discuss the evaluation results.

\section{Related work}

One approach to detect credible accounts uses heuristics such as the assumption that online opinion leaders are credible \cite{alsharawneh2013credibility}. This assumption is violated by misinformation superspreaders \cite{Grinberg2019,yang2021twitterfb,DeVerna2022FIB} and leaves out potentially credible ordinary users. 
Therefore this paper focuses on network-based approaches that leverage accounts' connectivity in addition to heuristics.


\textit{PageRank} \cite{page1999pagerank} is a widely used centrality measure that assigns scores for nodes in a directed network by simulating a diffusion process through the network analogous to random surfing among web pages. \textit{Personalized PageRank} \cite{haveliwala2003topic} incorporates prior knowledge about the importance of some nodes  by constraining the random surfer to stay close to those nodes. Methods extending PageRank and Personalized PageRank have been applied to measure trust in peer-to-peer (P2P) networks. \textit{EigenTrust} was used to obtain global reputation values for each user \cite{Eigentrust}; \textit{PowerTrust} \cite{PowerTrust} and \textit{TrustRank} \cite{trustrank,GuojunWANG:181} were used to rank search results. 
Different from existing work, one of our proposed methods attempts to infer account credibility by finding the highest-ranking nodes in a network where misinformation, rather than trust, propagates. Methods have been introduced that similarly model the spread of  ``distrust'' to measure trustworthiness in P2P networks \cite{Akavipat2009,PolarityRank,guha}. To our knowledge, no prior work has explored distrust in social news-sharing networks. 


Another well-known network centrality method is Hyperlink-Induced Topic Search (\textit{HITS}) \cite{kleinberg1999authoritative}. This method ranks the web pages returned by a search query by assuming that hubs are useful in leading a web surfer to authoritative pages. 
Several algorithms extend HITS. \textit{Co-HITS} \cite{deng2009cohitsgeneralized} and \textit{Bipartite Graph Reinforcement Model (BGRM)} \cite{rui2007bgrm} incorporate pre-existing information about the relevance of some web pages to constrain the final scores. \textit{BiRank} \cite{he2016birank} is a similar extension developed for \textit{n-}partite graphs.
In the context of social media, HITS has been used to find influential users \cite{Romero2011} as well as high-quality content \cite{agichtein2008finding}. 
One of our proposed methods extends HITS while maximizing prior knowledge about accounts to produce accurate credibility scores. The intuition is that misinformation sharers are hubs leading to unreliable news sources, and vice versa.


Machine learning methods have been employed to classify social media accounts based on their credibility. Previous studies have trained models on features engineered from a user profile \cite{castillo2013predicting,gupta2014tweetcred} or the content they post \cite{setiawan2020measuring,barbier2011information}. 
Graph embedding methods are popular means to obtain a compact representation of nodes in a network. Since network structure information is preserved, nodes with similar positions in the network have vectors that are close together in the embedding space \cite{node2vec,deepwalk,tang2015line}. These methods have been applied to classify rumor-spreading accounts in retweet networks, in combination with additional features such as an account's inferred believability \cite{rath2018utilizing}, screen name, profile description, and activity level \cite{hamdi2020hybrid}. 
We explore simpler methods that only use vectors capturing an account's position in the network in some of the algorithms proposed here.

\section{Methods}

The task is formalized as follows: given (i)~a set of posts with links to news articles and (ii)~credibility labels for a subset of accounts, assign credibility scores to the unlabeled accounts.

An overview of the pipeline for mining social media for this task is presented in Fig.~\ref{fig:framework}. The pipeline consists of five steps:

\begin{figure*}
    \centering
    \includegraphics[width=\fullwidth{}]{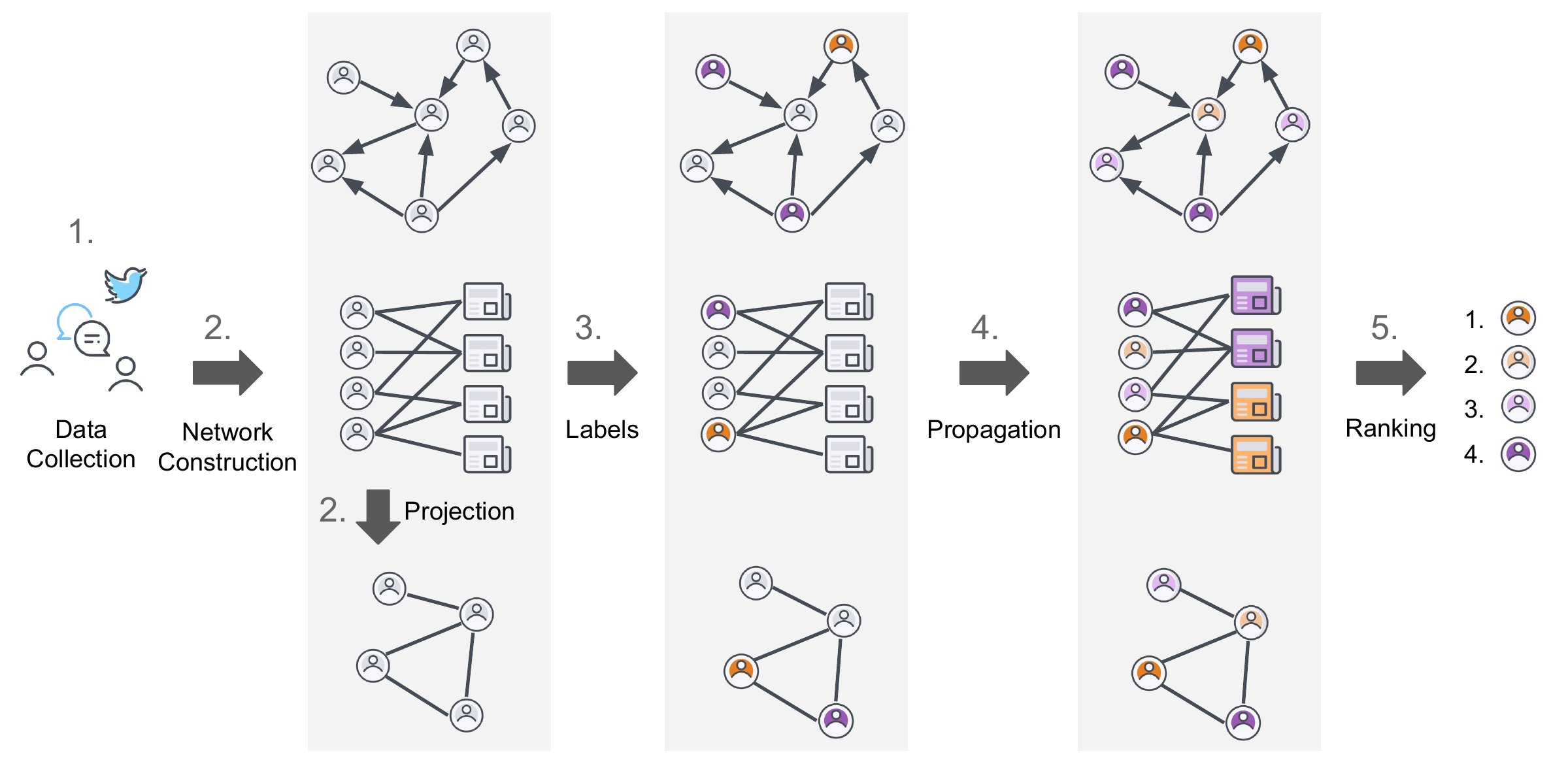}
    \caption{Pipeline for mining trust patterns from information-sharing data: trust in accounts is modeled by the reshare network (top); trust in sources can be modeled by the bipartite account-source network (middle), or a projection of it, the co-share networks (bottom). Colors represent credibility labels: orange and purple for high- and low-credibility, respectively.} 
    \label{fig:framework}
\end{figure*}

\begin{enumerate}
   \item Social media data is collected and cleaned. The data includes a set of known labels and a list of interactions between accounts, for example, via tweets or retweets.
   
   \item Different networks can be constructed to capture distinct potential signals. The reshare network captures an account's trust in other accounts. Some of our algorithms use the trust network --- the transpose of this reshare network.
   The bipartite account-source network captures an account's trust in sources. 
   The co-share network is obtained by projecting the bipartite network onto account nodes. 
    
   \item We label a subset of account nodes in the network. These labels are derived from source credibility scores provided by a fact-checking organization, such as NewsGuard\footnote{\url{newsguardtech.com}} or Media Bias Fact Check.\footnote{\url{mediabiasfactcheck.com}} 
   To obtain an account label, we calculate the weighted mean of all the sources they shared and then apply a threshold to this mean score, following the organization's standard for the credibility threshold.  
   
   \item Each algorithm is applied to propagate or compute credibility scores for all nodes.
   
   \item Lastly, unlabeled accounts are ranked by their credibility scores for evaluation.
\end{enumerate}

In the next subsections, we present several methods to infer account credibility based on trust in accounts or sources. In each case, we evaluate centrality-based methods and an embedding algorithm. Centrality-based methods are less sophisticated but more efficient and interpretable.

\subsection{Trust in accounts}

We describe algorithms to calculate an account's credibility by leveraging their trust in other accounts. Trustworthiness can be inferred from the \textit{trust network} $G^T$, in which a link goes from $Alice \rightarrow Bob$ if Alice follows or reshares Bob. In line with previous work, these actions can be considered endorsements that signal trust by the sharing account in the account being shared \cite{rath2018utilizing,roy,ZHAO20161,adali}.

The trust network is the transpose of the \textit{reshare network} $G$, a weighted, directed graph where the nodes represent accounts and edges correspond to reshare interactions among accounts. Edges follow the direction of information spread (from reshared to resharing account) and are weighted by the number of reshares. Let $G_{ij}=n$ if $j$ retweets $i$ $n$ times. 

Finally, we assume that some accounts have credibility labels. Let $H$ be a set of nodes that are known to have high credibility and $L$ a set of nodes known to have low credibility. 

\subsubsection{PageRank Trust}

The PageRank family of algorithms can be used to calculate account trustworthiness scores based on this signal. 
An account's \textit{PageRank Trust} score is calculated iteratively using a weighted version of PageRank:
\begin{equation}
    \tau_i^{t+1} = (1 - \alpha) \sum_j \frac{G^T_{ji}}{\sum_{\ell} G^T_{j \ell}} \tau_j^t + \frac{\alpha}{N} 
    \label{eq:PR} 
    \end{equation}
where $t$ is the iteration step, $\alpha$ is the teleportation factor. 
The intuition of this method is that accounts with many incoming links are trusted and therefore have high PageRank Trust scores, indicating high credibility.

\subsubsection{Personalized PageRank Trust}

Information about some high-credibility nodes may be available. To incorporate this information, an account's Personalized PageRank Trust is calculated as follows:
\begin{equation}
    \tau_i^{t+1} = (1 - \alpha) \sum_j \frac{G^T_{ji}}{\sum_{\ell} G^T_{j \ell}} \tau_j^t + \alpha \tau_i^0
    \label{eq:PPR} 
\end{equation}
where $\tau_i^0$ is defined by:
\begin{equation}
    \tau^0_{i} = \begin{cases}
       \frac{1}{|H|} & \text{if } i \in H \\
       0 & \text{otherwise.}
    \end{cases}
    \label{eq:init_ppr} 
\end{equation}
Under this scheme, accounts that are trusted by credible accounts ($H$ list) have higher Personalized PageRank Trust, indicating high credibility. 

\subsubsection{TrustRank}

This method first creates a quality seed set, ideally to be evaluated by experts, then uses this list to propagate scores. To apply the method in our context, PageRank Trust (Eq.~\ref{eq:PR}) is first used to select a set $S$ of seed accounts with highest PageRank Trust scores, where $|S|$ is a parameter. These seeds are labeled as ``good'' or ``bad'' depending on available credibility labels:
    \begin{equation}
        \tau_{i}^0= \begin{cases}
        1 & \text{if } i \in H \cap S\\
        0 & \text{if } i \in L \cap S\\
        \frac{1}{2} & \text{otherwise.}
        \end{cases}
    \end{equation}
TrustRank scores are then calculated using Personalized PageRank Trust (Eq.~\ref{eq:PPR}) with these $\tau^0_{i}$ values.

\subsubsection{LoCred}

The methods mentioned so far rely on the observation that good pages seldom point to bad ones \cite{trustrank}. Decisions to circulate news online are not as straightforward --- bad content is designed to mislead and people could be socially motivated. Therefore, we cannot assume that ``good people seldom trust bad ones,'' or be sure that credible accounts do not reshare from low-credibility accounts. 

We propose a method that uses the diffusion of information to capture an account's credibility without the assumption that accounts have good judgment. This method is performed on the reshare network. 
If a node reshares a lot, it might be gullible. These accounts might not spread misinformation intentionally but should be distrusted nonetheless. The \underline{Lo}w \underline{Cred}ibility Account Estimation (\textit{LoCred}) score is devised to capture such spreaders of misinformation.

The LoCred score $s_i$ for each node $i$ is calculated iteratively until convergence as a weighted version of personalized PageRank: 
\begin{equation}
    s_i^{t+1} = (1 - \alpha) \sum_j \frac{G_{ji}}{\sum_{\ell} G_{j \ell}} s_j^t + \alpha s_i^0
    \label{eq:locred} 
\end{equation}
where $t$ is the iteration step, $\alpha$ is the teleportation factor, and the initial value $s_i^0$ is defined as follows:
    \begin{equation}
        s^0_{i} = \begin{cases}
        \frac{1}{|L|} & \text{if } i \in L \\
        0 & \text{otherwise.}
        \end{cases}
        \label{eq:init_locred} 
    \end{equation}
Accounts close to nodes in $L$ have higher LoCred scores, indicating low credibility.
Note that the initial values in Eqs.~\ref{eq:init_ppr} and \ref{eq:init_locred} are different. But even if they were the same, the rankings based on Eqs.~\ref{eq:PPR} and \ref{eq:locred} would not simply be the reverse of each other because the eigenvectors of a matrix and those of its transpose are different for asymmetric matrices.

\subsubsection{Reputation Scaling}

Lastly, we introduce \textit{Reputation Scaling}, a measure that balances the perceived trustworthiness of an account and its spreading of misinformation. This method builds on an application for distributed P2P systems \cite{Akavipat2009}. 

The reputation of an account is ``scaled'' by combining both its LoCred and Personalized PageRank Trust score.
We calculate both of these scores, then rank accounts based on a reputation score calculated as follows:
    \begin{equation}
        r_i = \tau_i (1 - s_i)
    \end{equation}
where $\tau_i$ is the trust of $i$ calculated with Eq.\ref{eq:PPR} and $s_i$ is its LoCred score calculated with Eq.\ref{eq:locred}. 
An account with high reputation is considered credible. 

\subsubsection{Node2vec on reshare network}

We investigate whether an account's network position provides predictive signals about its credibility, i.e., accounts with similar network positions are similarly credible. To this end, we use node2vec \cite{node2vec} to obtain embedding vectors of accounts from the retweet network $G$. 
We use the typical values for node2vec parameters. Specifically, vector dimension, window size, number of walks, and walk length are 128, 10, 10, and 80, respectively. The optimization is run for ten epochs. 
Account node2vec vectors are then used in a $k$-Nearest Neighbors (KNN) classifier \cite{altman1992knn} with $k=10$ to rank unlabeled accounts. 
We find the best results by performing a grid search over the random walk bias parameters $p, q \in \{0.25, 0.50, 1, 2, 4\}$. We evaluate each parameter combination using 5-fold cross-validation with 20\% labeled data in each fold.

\subsection{Trust in sources}

We describe algorithms to calculate the credibility of accounts by leveraging their trust in sources (websites or domains in our case). We hypothesize that the mutually reinforcing relationship between news websites and news consumers can be applied in our bipartite account-source network to infer account credibility. Consider accounts as hubs and news domains as authorities: a ``good'' account is one pointing to many ``good'' ---authoritative, high-credibility--- websites; and vice versa, a good website is a node shared by many good accounts \cite{kleinberg1999authoritative}.

Let us define the \textit{account-source network} as a weighted bipartite graph where nodes consist of two disjoint sets $U$ and $D$ that represent accounts and sources, respectively. A weighted edge represents the number of times an account shares links to a source. Formally, let $G$ be the adjacency matrix such that $G_{ij}=n$ if account $i \in U$ shares links from source $j \in D$ $n$ times. 
Similar to previous algorithms, we assume that some accounts have credibility labels.

\subsubsection{HITS}

HITS \cite{kleinberg1999authoritative} updates account and source scores according to 
\begin{align}
    u_i^{t+1} &= \sum_{j \in D} A_{ij} d_j^{t} \\
    d_j^{t+1} &= \sum_{i \in U} A_{ij} u_i^{t}
\label{eq:hits}
\end{align}
respectively, where $A_{ij} = 1 \text{ if } G_{ij} > 0$ and zero otherwise. The scores are normalized after each step $t$.

\subsubsection{Co-HITS}

Information about some high-credibility sources may be available. Co-HITS incorporates this information into the propagation to calculate node scores \cite{deng2009cohitsgeneralized}. Account and source scores are calculated iteratively according to the strength of interactions between nodes from the two network partitions, until convergence:
\begin{align}
    u_i^{t+1} = \alpha u_i^0 + (1-\alpha)\sum_{j} \frac{G_{ij} d_j^{t}}{\sum_\ell{G_{i \ell}}} \\
    d_j^{t+1} = \beta d^0 + (1-\beta)\sum_{i} \frac{G_{ij} u_i^{t}}{\sum_\ell{G_{\ell j}}} 
    \label{eq:cohits:d}
\end{align}
respectively, where $\alpha$ and $\beta$ are the teleportation factors for accounts and sources respectively, $d^0 = 1/|D|$ is the initial value for sources, and $u_i^0$ is the initial value for account $i$, defined as: 
\begin{equation}
     u^0_{i} = 
    \begin{cases}
    0 & \text{if } i \in H \\
    1 & \text{if } i \in L \\
    1/|U| & \text{otherwise.}
    \end{cases}
    \label{eq:cohits_u_init}
\end{equation}
These initial values are further normalized such that $\sum_{i \in U} u^0_i = 1$.
The scores are normalized, so no further normalization is necessary. 

\subsubsection{BGRM}
BGRM \cite{rui2007bgrm} is similar to Co-HITS. The main difference is in the way they normalize node scores at each iteration (see Table~1 in \cite{he2016birank} for more details).

\subsubsection{BiRank}
BiRank \cite{he2016birank} is also similar to Co-HITS and BGRM, differing in the way node scores are normalized at each iteration (see Table~1 in \cite{he2016birank} for more details). 

\subsubsection{CoCred}

We propose \underline{Co}-sharing network-based \underline{Cred}ibility (\textit{CoCred}), a measure that, like Co-HITS, BGRM, and BiRank, incorporates existing knowledge about some accounts' credibility into the credibility estimation of other accounts. 
The difference is that those algorithms apply the update rules to labeled accounts, whereas CoCred preserves the known labels, only rescaling them in the normalization step --- we believe the accounts labels contain very strong signals.

The CoCred score $u_i$ for each account $i$ and the corresponding score $d_j$ for each source $j$ are updated at each timestep $t$ according to
\begin{align}
    u_i^{t+1} &= 
    \begin{cases}
    u_i^0 & \text{if } i \in H \cup L \\
    \alpha u_i^0 + (1-\alpha)\sum_{j} \frac{G_{ij} d_j^{t}}{\sum_\ell{G_{i \ell}}} &  \text{otherwise}
    \label{eq:cocred_u}
    \end{cases}
    \\
    d_j^{t+1} &= \beta d^0 + (1-\beta)\sum_{i}\frac{G_{ij} u_i^{t}}{\sum_\ell{G_{\ell j}}}
    \label{eq:cocred_d}
\end{align}
where $\alpha$ and $\beta$ are the teleportation factors for accounts and sources, respectively; $d^0 = 1/|D|$ is the initial value for sources, and $u_i^0$ is the initial value for account $i$, as defined in Eq.~\ref{eq:cohits_u_init}. Note that Eq.~\ref{eq:cocred_d} is the same as Eq.~\ref{eq:cohits:d}.

After each update, the scores are normalized so that $\sum_{i \in U} u_i = \sum_{j \in D} d_j = 1 $.
Our algorithm considers low-credibility accounts as those sharing unreliable news sources and vice versa. Low-credibility accounts have higher CoCred scores.

\subsubsection{Node2vec on co-share network}

The bipartite account-source network may also provide information about similar news-sharing patterns between accounts. To explore this, we project the bipartite network onto account nodes. 
Specifically, the \textit{co-share network} is obtained by connecting accounts if they share links to the same sources.
It is thus a weighted, undirected graph where the nodes are accounts having at least one shared source in common with another account.
Each account is represented as a vector of shared domains. 
To mitigate the effect of popular sources on the similarity among accounts, we use Term Frequency - Inverse Document Frequency (TF-IDF) vector representations \cite{jones1972tfidf} where each dimension is a news source. 
Edges correspond to co-share interactions and are weighted by the cosine similarity between the TF-IDF vectors of the two connected accounts.
Account embedding vectors obtained from the co-share network may reveal whether individuals who share information from the same sources have similar credibility. 
We use node2vec to obtain account vectors and use them in a KNN classifier with $k=10$ to rank unlabeled accounts. The evaluation of this method uses the same parameter set and optimization procedure as that of node2vec on the reshare network described above.

\section{Evaluation}

Our evaluation framework only requires a set of social media posts, and not follower/friend data. 
We define the task of classifying low-credibility accounts as follows: given (i) a set of social media posts with links to news articles, and (ii) credibility labels for a subset of accounts, assign a binary label --- \emph{credible} or \emph{not credible} --- to the unlabeled accounts. 
The algorithms are evaluated on networks extracted from social media data. The datasets, corresponding networks, experimental setup, and results are described below.

\subsection{Data}

We consider three social media datasets.

\begin{itemize}
    \item The \coviddataset{} dataset includes tweets about COVID-19, collected with the hashtags \texttt{\#coronavirus} and \texttt{\#covid19} from 9--29 March 2020 \cite{yang2020prevalence}. 
    \item The \midtermdataset{} dataset includes tweets containing hashtags and keywords about the U.S. 2022 Midterm elections \cite{aiyappa2023multi}. We use a subset of this collection, from 8 October--18 November 2022. 
    \item The \fbdataset{} dataset contains Facebook posts collected the same way and spanning the same period as \midtermdataset{} \cite{aiyappa2023multi}. 
\end{itemize}

We extract the data fields of interest from each social media post, including the user ID, post ID, and domain names of the shared links. If a link is shortened when shared on these platforms, a pre-processing step is performed to extract the domain name from the expanded URL. Posts with domain names of platforms such as \url{amazon.com}, \url{yelp.com}, \url{youtube.com}, etc. are not considered. To limit noise, we further retain only accounts sharing at least five links and posts containing domains that are shared at least five times across the dataset. 
To assign scores to accounts, we start from source credibility scores, which are domain ratings obtained from NewsGuard in April 2021 for the \coviddataset{} and October 2022 for the \midtermdataset{} and \fbdataset{}.
We then compute an account score as a weighted mean of the scores of the sources they share. Note that not all accounts have a score as a result. Finally, following NewsGuard's rubric, we use a threshold of 60 for labeling accounts with a score below/above 60 as low/high-credibility. 

\subsection{Network description}

\subsubsection{Reshare network}

Since reshare information is not included in Facebook data, the reshare network can only be constructed from the two Twitter\footnote{Because our data was collected from Twitter before it was renamed `X,' we continue to refer to the platform by its original name.} datasets out of the three datasets considered. The statistics of these retweet networks are summarized in Table~\ref{table:rt_stats}. The table reports on network assortativity coefficients \cite{newman2003mixing} using account credibility scores. The high coefficients are indicative of credibility homophily (answering \textbf{Q1}). 


\begin{table}
\centering
\caption{Retweet network descriptions.}
\begin{tabular}{lrrrr}
\hline
Dataset & Nodes & Edges & Avg. degree & Credibility assortativity coeff.\\  
\hline
\coviddataset{} &  322,208 & 382,499 &  1.2 & 0.83\\ 
\midtermdataset{} & 410,914 & 762,534 & 1.9 & 0.73\\
\hline
\end{tabular}
\label{table:rt_stats}
\end{table}

\begin{figure}
    \centering
    \includegraphics[width=\fullwidth{}]{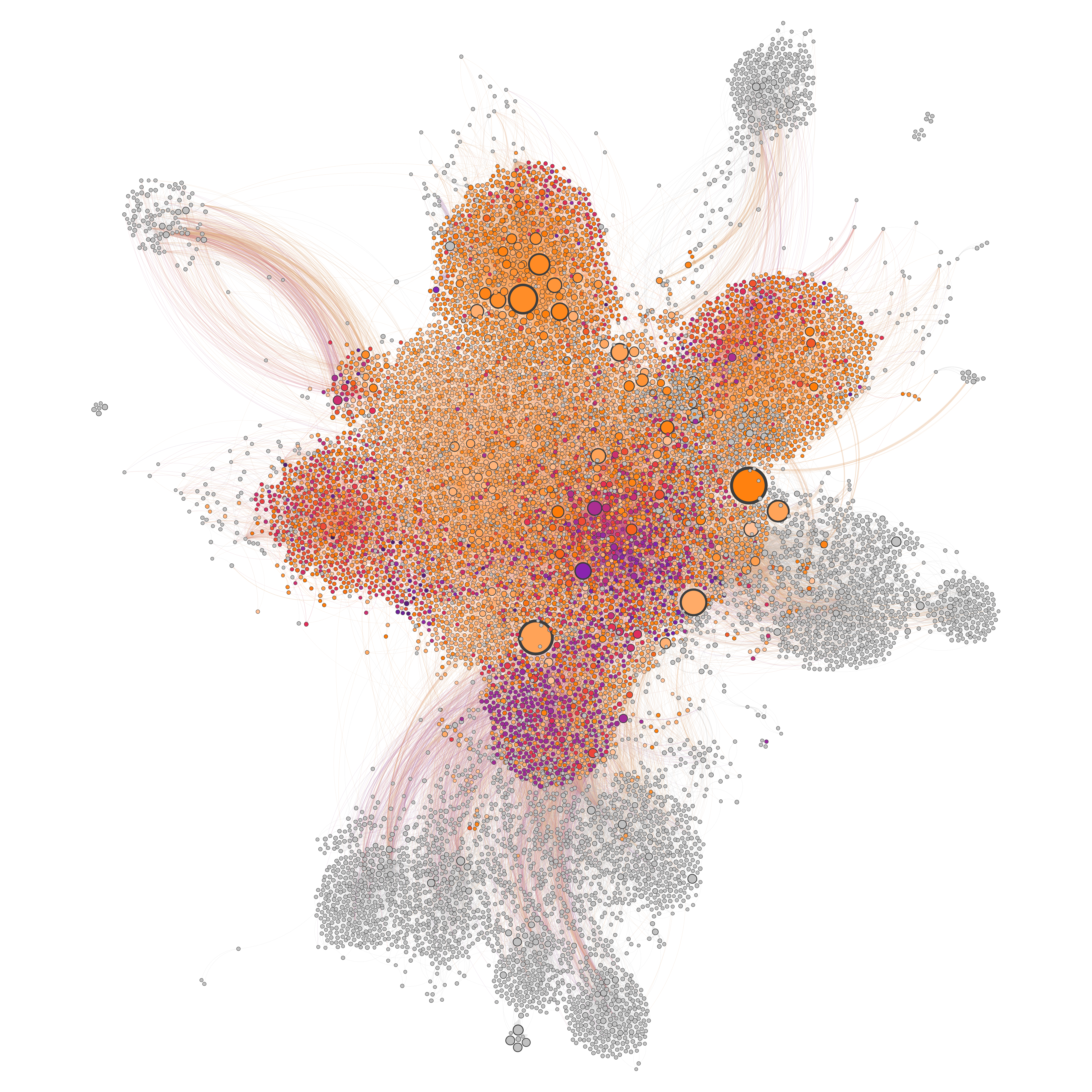}
    \caption{Retweet network constructed from the \coviddataset{} dataset. To visualize the network, we filter nodes and edges using k-core decomposition \cite{alvarez2006large} ($k=3$). Colors represent credibility score: orange and purple for high- and low-credibility accounts, respectively, and gray for unlabeled accounts. Node size represents in-strength, i.e., the number of retweets by the account. Large purple nodes are known misinformation super-spreaders.}
    \label{fig:retweet_network}
\end{figure}

The homophily can also be observed in the core of the retweet network (Fig.~\ref{fig:retweet_network}), where low-credibility accounts tend to reshare from low-credibility accounts, and vice versa. 

\subsubsection{Bipartite network}


\begin{table}
\centering
\caption{Bipartite network descriptions.}
\begin{tabular}{lrrrr}
\hline
Dataset & Accounts & Domains & Edges & Avg. degree \\  
\hline
\coviddataset{} & 474,094 & 55,539 & 613,609 & 1.3\\
\midtermdataset{} & 126,445 & 13,415 & 1,190,390 & 9.4\\ 
\fbdataset{} & 6,950 & 4,141 & 26,252 & 17.1\\ 
\hline
\end{tabular}
\label{table:bipartite_stats}
\end{table}

\begin{figure}
    \centering
    \includegraphics[width=\fullwidth{}]{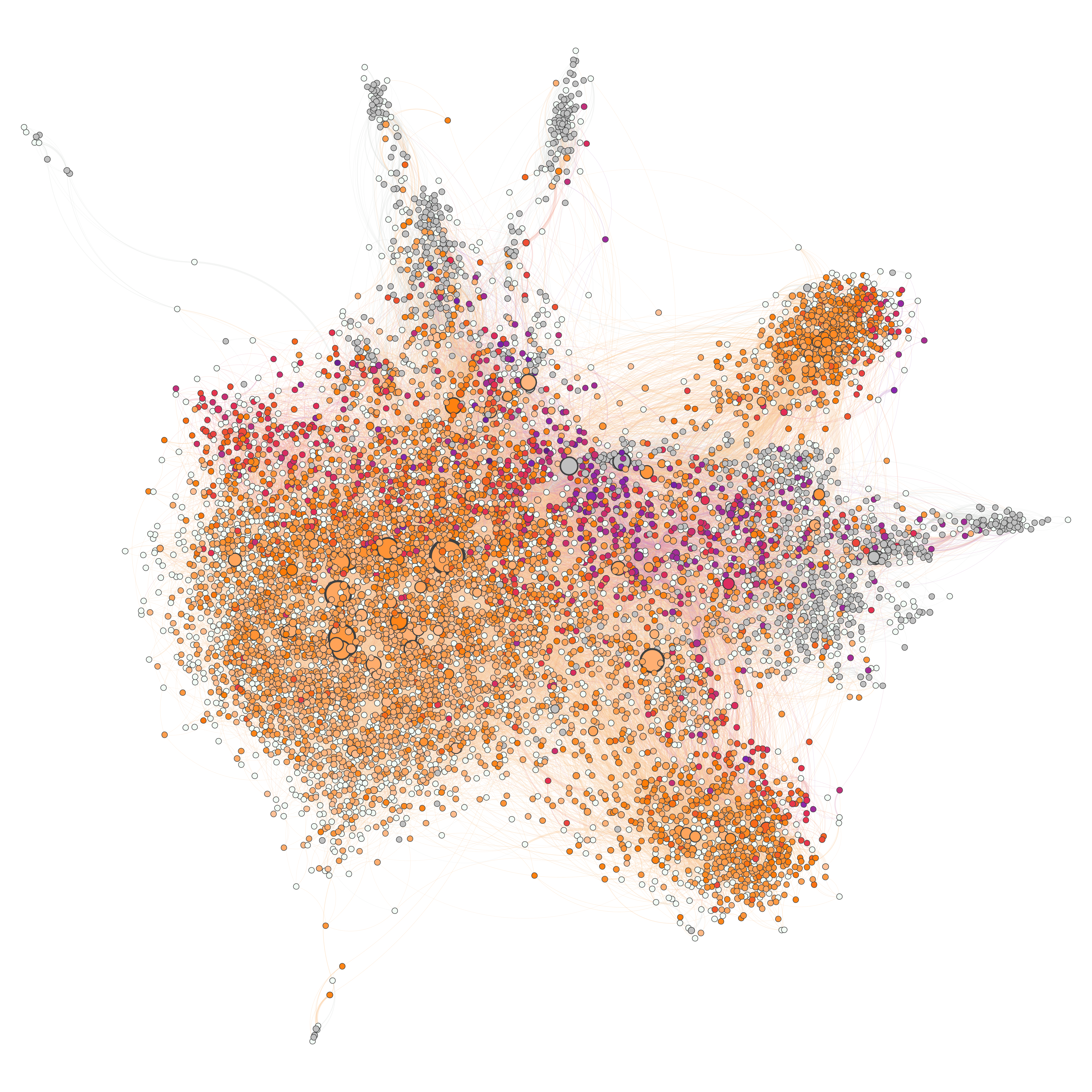}
    \caption{Bipartite account-source network constructed from the \coviddataset{}. For visualization, only the $k=4$ core of the network is shown. Node colors represent account credibility scores: orange and purple for high- and low-credibility accounts, respectively, and gray for unlabeled accounts. Source nodes are in white. Node size represents the strength of an account node, i.e., the number of posts with links by that account. 
    }
    \label{fig:bipartite_network}
\end{figure}

The statistics of the bipartite networks are summarized in Table~\ref{table:bipartite_stats}.
The core of a bipartite network is visualized in Fig.~\ref{fig:bipartite_network}.

\subsubsection{Co-share network}

\begin{table}
\centering
\caption{Co-share network descriptions.}
\begin{tabular}{lrrrr}
\hline
Dataset & Nodes & Edges & Avg. degree & Credibility assortativity coeff.\\   
\hline
\coviddataset{} &  67,657 & 434,963 &  1,441.7 & 0.5 \\
\midtermdataset{} & 115,461 & 59,491,099 & 515.3 & 0.8 \\
\fbdataset{} & 3,488 & 59653 & 34.2 & 0.7\\ 
\hline
\end{tabular}
\label{table:cosharing_stats}
\end{table}

\begin{figure}
    \centering
    \includegraphics[width=\fullwidth{}]{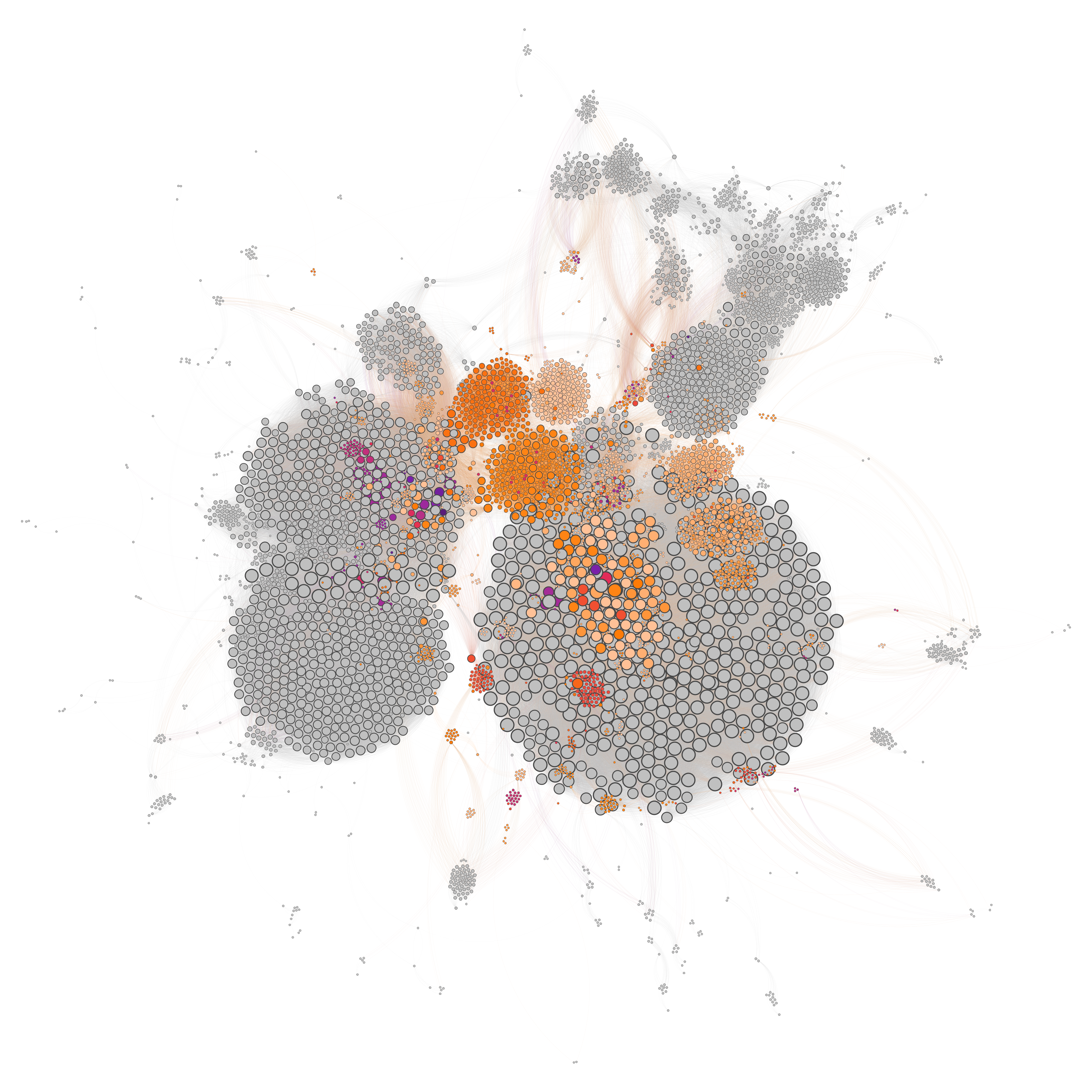}
    \caption{Co-share network constructed from the \coviddataset{}. Given the density of this network, we aggressively filter it for visualization. We first keep the 10\% of edges with the highest weights and the 10\% of nodes in the innermost core ($k=4$). Then we apply the multiscale backbone method \cite{serrano2009extracting} (significance level 0.3) and retain the giant component of the backbone network. Colors represent credibility score: orange and purple for high- and low-credibility accounts, respectively; unlabeled accounts are in gray. Node sizes represent core numbers: smaller nodes are in the periphery of the network.}
    \label{fig:coshare_network}
\end{figure}

Descriptive statistics of the co-share networks are presented in Table~\ref{table:cosharing_stats}. There is an assortative mixing of credibility among accounts, suggesting the existence of credibility homophily in the co-share networks (answering \textbf{Q2}). 
The core backbone of the co-share network, shown in Fig.~\ref{fig:coshare_network}, illustrates this other type of homophily: similarly credible accounts tend to share content from similar sources. 

\subsection{Experimental setup}

We employ the area under the receiver operating characteristic curve (ROC\_AUC) and F1 score as metrics for low-credibility account classification. 
The maximum value of ROC\_AUC, one, indicates that all low-credibility accounts are ranked before other accounts, whereas a value of 0.5 corresponds to random ranking. F1 is the harmonic mean of precision and recall. An F1 value of one means that the model correctly classifies all classes; zero means all samples are wrongly classified. 
The F1 score requires converting the account credibility scores provided by the algorithms into binary labels using a threshold. We calculate precision and recall for a thousand threshold values spanning the unit interval and select the optimal threshold that maximizes the F1 score. 

Since NewsGuard ratings are available for only a subset of domains shared in a dataset, we assign a \emph{label confidence score} to an account by calculating the fraction of known domains that they shared; this number is between zero and one. The most stringent threshold is used in our evaluation --- only accounts with a confidence score of one are used as \emph{known accounts} to calculate evaluation metrics. 

The evaluation uses 5-fold cross-validation. We hold out a random subset of 20\% of known accounts (the test set). For centrality-based methods on the reshare and bipartite network, the initial scores of the remaining 80\% of accounts are set to their true scores, while the test set starts with default scores at the beginning of propagation. Similarly, for classification using embeddings, we train the classifier on the remaining 80\% of accounts. Finally, for both cases, predicted labels for accounts in the test set are checked against the true labels. The reported metrics are averages of values across five folds. 

We use $\alpha=\beta=0.85$ for the teleportation factors in all network centrality algorithms. 
The best node2vec parameters for the retweet networks are $p=q=1.0$.
For the co-share network, the best node2vec parameters are $p=2.0$, $q=0.5$ for the \coviddataset{} dataset, $p=q=1.0$ for the \midtermdataset{} dataset, and $p=0.25$, $q=4.0$ for the \fbdataset{} dataset.

\subsection{Results}

Evaluation results for the low-credibility account classification task are summarized in Tables~\ref{table:results_auc} (based on ROC\_AUC) and \ref{table:results_f1} (based on F1 scores). The trends in performance are similar for both metrics, therefore we focus on ROC\_AUC.


\begin{table}
\centering
\caption{Account classification ROC\_AUC. Best results for each dataset are highlighted in darker shades.}
\begin{tabular}{llcccccc}
 \hline
Sharing network & Method & \coviddataset{} & \midtermdataset{} & \fbdataset{}\\ 
\hline
 \multirow{7}*{Reshare} & node2vec & \colorbest{}$0.910\pm0.006$ & \colorbest{}$0.885\pm0.003$ & \multirow{7}*{N/A} \\
 & LoCred & $0.768\pm0.004$ & $0.773\pm0.044$ &\\  
& Rep. Scaling & $0.618\pm0.008$ & $0.660\pm0.003$ \\ 
& Trustrank & $0.534\pm0.002$ & $0.517\pm0.001$ \\
& PPR Trust & $0.534\pm0.002$ & $0.517\pm0.001$ \\
& PR Trust & $0.516\pm0.002$ & $0.520\pm0.002$\\  
 
\hline
\multirow{4}*{Bipartite/Co-share} 
& CoCred & \colorsecondbest{}$0.802\pm0.014$ & $0.831\pm0.013$ & \colorsecondbest{} $0.715\pm0.042$\\  
& node2vec & $0.654\pm0.020$ & \colorsecondbest{}$0.873\pm0.005$ & \colorbest{}$0.829\pm0.056$\\
& Co-HITS & $0.699\pm0.047$ & $0.733\pm0.018$ & $0.707\pm0.062$\\  
& HITS & $0.673\pm0.003$ & $0.520\pm0.006$ & $0.626\pm0.043$\\
& BGRM & $0.559\pm0.003$ & $0.513\pm0.001$ & $0.614\pm0.042$\\
& BiRank & $0.544\pm0.003$ & $0.504\pm0.008$ & $0.576\pm0.031$\\
\hline
\end{tabular}
\label{table:results_auc}
\end{table}

The results show that the trust relationships among accounts provide useful information to effectively infer their credibility. 
In particular, the best-performing method is node2vec on reshare networks, with a ROC\_AUC score of more than 0.88 across datasets.
The next best-performing algorithms are LoCred and Reputation Scaling; these algorithms consider reshares as information diffusion channels without assuming good judgment by accounts. TrustRank, Personalized PageRank Trust, and PageRank Trust, which assume that reshares capture trust and hence reputation, perform only marginally better than random at the task.
These results reinforce the difference between web-surfing behavior and news-sharing behavior on social media, where the assumption that good accounts seldom trust bad ones might not hold.  


The bipartite/co-share networks capturing account trust in sources also provide effective information to estimate account credibility. 
The best-performing algorithm in each dataset achieves a ROC\_AUC score of more than 0.8 (CoCred in the \coviddataset{} dataset and node2vec in the \midtermdataset{} and \fbdataset{} datasets). 
Among the centrality-based algorithms on the bipartite network, our proposed algorithm CoCred performs the best, followed by Co-HITS and HITS; BGRM and BiRank perform close to random. 
CoCred's high ROC\_AUC score confirms our intuition that the known account labels provide strong signals and should be preserved in score propagation to produce accurate ratings. 

\begin{table}
\centering
\caption{Account classification F1. Best results for each dataset are highlighted in darker shades.}
\begin{tabular}{llcccccc}
 \hline
Sharing network & Method & \coviddataset{} & \midtermdataset{} & \fbdataset{}\\ 
\hline
 \multirow{7}*{Reshare} & node2vec & \colorbest{}$0.918\pm0.003$ & \colorbest{}$0.886\pm0.003$ & \multirow{7}*{N/A} \\
& LoCred & $0.786\pm0.005$ & $0.757\pm0.034$\\  
& Rep. Scaling & $0.576\pm0.006$ & $0.514\pm0.007$\\ 
& Trustrank & $0.196\pm0.004$ & $0.214 \pm0.002$\\
& PPR Trust & $0.196\pm0.004$ & $0.218\pm0.002$ \\
& PR Trust & $0.176\pm0.012$ & $0.251\pm0.002$\\  
 
\hline
\multirow{4}*{Bipartite/Co-share} 
& CoCred & \colorsecondbest{}$0.800\pm0.022$ & $0.778\pm0.014$ & \colorsecondbest{}$0.756\pm0.032$\\  
& node2vec & $0.707\pm0.024$ & \colorsecondbest{}$0.886\pm0.008$ & \colorbest{}$0.827\pm0.029$\\
& Co-HITS & $0.638\pm0.016$ & $0.655\pm0.019$ & $0.728\pm0.051$\\  
& HITS & $0.738\pm0.004$ & $0.199\pm0.010$ & $0.628\pm0.066$\\
& BGRM & $0.501\pm0.002$ & $0.146\pm0.002$ & $0.411\pm0.121$\\
& BiRank & $0.477\pm0.011$ & $0.163\pm0.089$ & $0.434\pm0.106$ \\
\hline
\end{tabular}
\label{table:results_f1}
\end{table}                          

In summary, account credibility can be estimated accurately using trust signals from news-sharing networks.
While all networks provide useful signals, the reshare network is slightly more informative for the task than the bipartite/co-share networks, based on the best-performing algorithm in each dataset. 

\begin{figure}
    \centering
    \includegraphics[width=\fullwidth{}]{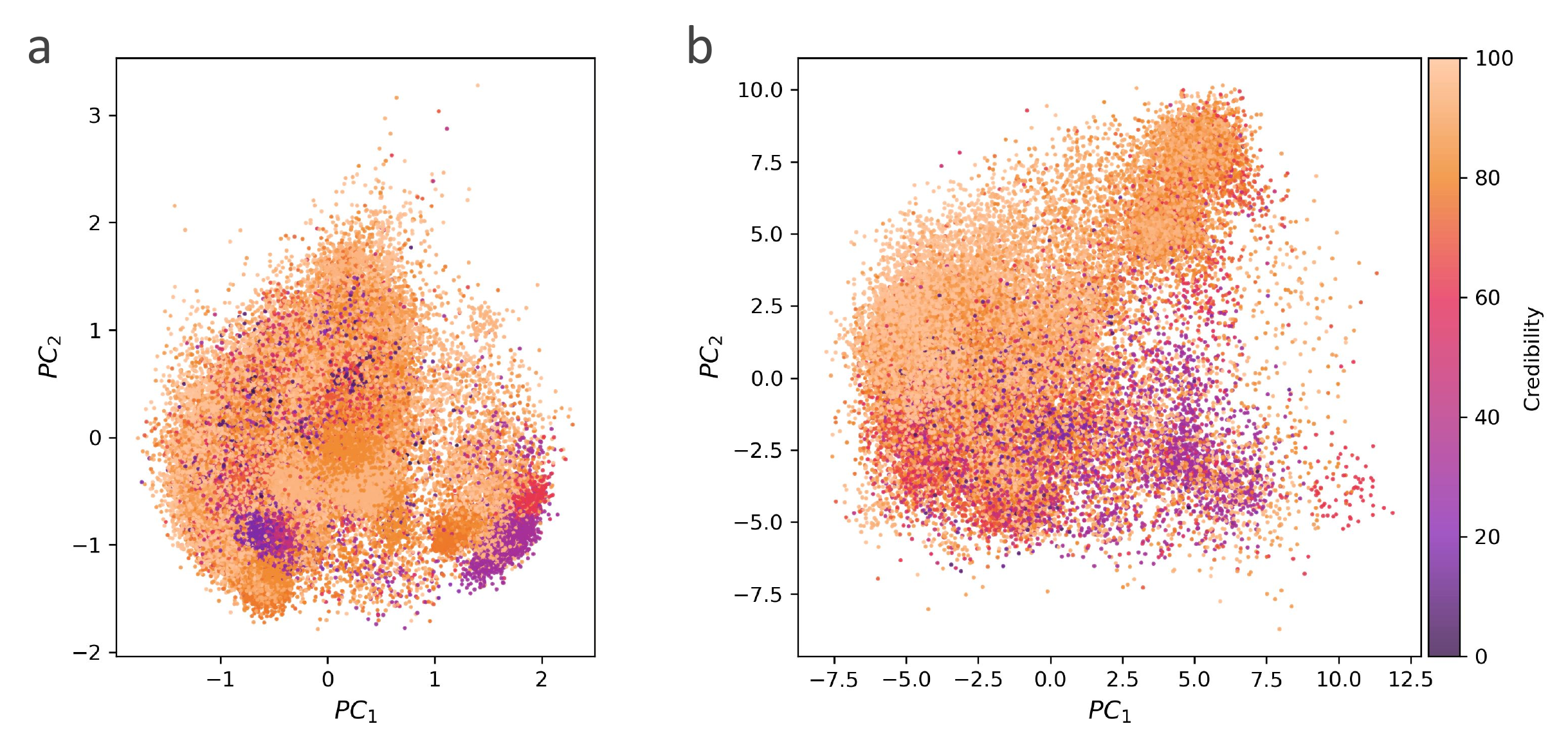}
    \caption{Account embedding vectors for the \coviddataset{} dataset. Node2vec vectors representing accounts are projected onto a two-dimensional space using PCA for (a)~reshare and (b)~co-share networks. Colors represent account credibility scores: orange and purple for high- and low-credibility accounts respectively.}
    \label{fig:pca}
\end{figure}

Fig.~\ref{fig:pca} shows node2vec embedding vectors of accounts, with their dimensionality reduced using Principle Component Analysis (PCA) \cite{labrin2020principal}. We observe that the embeddings in the reshare network more effectively cluster accounts with similar credibility, compared to the embeddings in the co-share network.

\section{Discussion}

This paper suggests ways to mitigate misinformation on social media by detecting low-credibility accounts likely to spread misinformation. To this end, we introduce methods to infer account credibility based on two information-sharing networks: a reshare network capturing account trust in other accounts and a bipartite network reflecting account trust in information sources. 
A systematic evaluation shows that the proposed methods are effective in detecting low-credibility accounts. 
The most successful method utilizes node2vec embeddings of accounts on reshare networks, with both ROC\_AUC and F1 around 0.9.  
Our proposed network centrality algorithms, LoCred and CoCred, consistently perform well across the three datasets considered, with ROC\_AUC and F1 both around 0.8.
We find two kinds of credibility homophily in the news-sharing networks, which help explain the effectiveness of these methods: unreliable accounts tend to reshare content from one another (\textbf{Q1}), and share content from similar sources (\textbf{Q2}).
These results confirm our hypothesis that the structure of the reshare or co-share network provides strong signals to effectively infer account credibility (\textbf{Q3}). While the presence of credibility homophily has also been observed in prior analyses of different Twitter data \cite{Nikolov2020partisanship}, future work might consider whether these results generalize to information-sharing networks derived from other social media platforms. 

While our framework detects potential misinformation spreaders regardless of intention, it can help to better characterize disinformation spreaders in combination with other effective methods, such as bot and inauthentic coordinated campaign detection \cite{yang2019arming,pacheco2021uncovering}.  
In particular, one of our proposed methods, CoCred, provides source rankings simultaneously with account rankings. These source rankings can be used to estimate the credibility of emerging news media outlets. Such source credibility scores can be used as a feature in misinformation classification, complementing extant content-based methods to curb misinformation.

Our work has some limitations. 
First, we only label and investigate accounts sharing sources with ratings by NewsGuard. Additional sources of misinformation that are not labeled by NewsGuard, as well as false/misleading claims in individual posts that do not include a link to some source, are missed in our analysis. However, since the criteria used by NewsGuard to label sources are independent from the methods we evaluate, there is no reason to believe that our accuracy measurements are biased. Still, complementary analyses using other sources of ground truth, such as post-level or account-level annotations, could provide additional insights. 
Second, classification results depend on the news-sharing data from which a network is derived. As such, account credibility scores may vary across contexts. 
For example, users have a higher tendency to share links to unconfirmed sources \cite{Oh2010rumor} when information is scarce, diverging from their usual news-sharing behavior. Accounts might be classified as unreliable in these situations, despite being reliable in others. 
One way to address this concern might be to aggregate an account's credibility scores across diffusion networks from different topics. 
Lastly, algorithms using the reshare network cannot be evaluated on the Facebook platform, as the data does not provide information about reshares. Furthermore, Twitter/X has recently removed free data access for researchers, making this kind of analysis difficult to replicate in the future. Note, however, that the proposed methods are platform-agnostic. The outlined analyses can be applied to data from any platform that allows for the construction of reshare or bipartite networks, such as Mastodon and Bluesky. We urge all social media platforms to make data available to allow for transparent analyses \cite{misinfo_data}.

Overall, this work enriches the understanding of misinformation spreaders on social media by investigating trust signals in different information-sharing networks.
The observed credibility homophily invites further exploration of misinformation diffusion and spreader dynamics. The proposed methods can be used to identify accounts of interest based on their estimated credibility by considering only sharing metadata, before further scrutiny (or even without any scrutiny) of the specific content they share. Platforms can readily apply our proposed methods to enhance their moderation efforts. 
These insights are crucial to social media platforms and policymakers in the debate on ways to combat online misinformation.

\begin{singlespace}
\section{Abbreviations}

\begin{tabbing}
BGRM ~~~~~ \=  Bipartite Graph Reinforcement Model\\
CoCred \> Co-sharing network-based Credibility \\
HITS \> Hyperlink-Induced Topic Search \\
KNN \> k-Nearest Neighbors \\
LoCred \> Low Credibility Account Estimation \\
P2P \> peer-to-peer \\
PCA \> Principle Component Analysis  \\
PPR Trust \> Personalized PageRank Trust \\
PR Trust \> PageRank Trust \\
ROC\_AUC \> receiver operating characteristic curve \\
TF-IDF \> Term Frequency - Inverse Document Frequency \\
\end{tabbing}

\section*{Declarations}

\subsection*{Ethics approval} 

This research is based on analyses of public social media data with minimal risks to participants. 
This study has been granted exemption from review by the Indiana University IRB (protocol 17036).
The collection and release of the dataset are in compliance with the platforms' terms of service.

\subsection*{Consent for publication} 

Not applicable.

\subsection*{Availability of data and material} 

For reproducibility, we make the data and code available in a public repository \url{https://github.com/osome-iu/credibility-inference}.

\subsection*{Competing interests}

The authors declare that they have no competing interests.

\subsection*{Funding}

This research is supported in part by the Knight Foundation, Craig Newmark Philanthropies, DARPA (awards W911NF-17-C-0094 and HR001121C0169), and the Luddy School of Informatics, Computing, and Engineering at Indiana University, Bloomington.

\subsection*{Authors' contributions}

FM and BT formulated the research, developed the methodology, and prepared the manuscript. BT collected the data. BT and OA performed analyses. All authors read and approved the final manuscript. 

\subsection*{Acknowledgements} 

We are grateful to Kai-Cheng Yang for help with the data collection and many helpful comments. Ruj Akavipat and Pik-Mai Hui provided helpful discussions. We also thank NewsGuard for licensing their source credibility scores.

\bibliographystyle{bmc-mathphys} 
\bibliography{bib}      

\end{singlespace}

\end{document}